\documentclass[prl,byrevtex,nobalancelastpage,showpacs,twocolumn,amsmath,amssymb]{revtex4}
\usepackage{graphicx}
\usepackage{color}

\newcommand{\oL}{\hat{L}_\lambda}               % Fokker-Planck operator
\newcommand{\oW}{\hat{L}_\lambda^W}             % work operator
\newcommand{\oP}{\hat{\Pi}_\lambda}             % projection operator
\newcommand{\oA}{\hat{A}_\lambda}               % short operator
\newcommand{\eq}{\text{eq}}                     % equilibrium
\newcommand{\ts}{t_\text{s}}                    % switching time
\newcommand{\feq}{f^\eq_\lambda}                % Boltzmann's distribution
\newcommand{\dlam}{\dot{\lambda}}
\newcommand{\dV}{\pd{V_\lambda}{\lambda}}
\newcommand{\dd}{\text{d}}                      % for integral/derivative
\newcommand{\pd}[2]{\frac{\partial #1}{\partial #2}}    % partial derivative
\newcommand{\mean}[1]{\left\langle #1 \right\rangle}    % average
\renewcommand{\vec}[1]{\mathbf #1}              % vector
\newcommand{\igl}[3]{\int_{#2}^{#3}\dd #1\;}    % integral
\newcommand{\ugl}[1]{\int\dd #1\;}              % integral

\begin{document}

%\today

\title{Distribution of work in isothermal non-equilibrium processes}
\author{Thomas Speck}
\author{Udo Seifert}
\affiliation{{II.} Institut f\"ur Theoretische Physik, Universit\"at Stuttgart,
70550 Stuttgart, Germany}
\pacs{05.40.-a, 05.70.Ln}

\begin{abstract}
Diffusive motion in an externally driven potential is considered. It is
shown that the distribution of work required to drive the system from an
initial equilibrium state to another is Gaussian for slow but finite
driving. Our result is obtained by projection method techniques exploiting a
small parameter defined as the switching rate between the two states of the
system. The exact solution for a simple model system shows that such an
expansion may fail in higher orders, since the mean and the variance
following from the exact distribution show non-analytic behavior.
\end{abstract}

\maketitle

%%%%%%%%%%%%%%%%%%%%%%%%%%%%%%%%%%%%%%%%%%%%%%%%%%%%%%%%%%%%
%% Introduction
%%%%%%%%%%%%%%%%%%%%%%%%%%%%%%%%%%%%%%%%%%%%%%%%%%%%%%%%%%%%

In macroscopic thermodynamics, the work $W$ spent in changing the state 
of a system at constant temperature $T$ obeys 
\begin{equation}
  W \geqslant \Delta F,
\end{equation}
which is one version of the second law where $\Delta F$ is the difference in 
free energy of the final and the initial equilibrium state. As the system gets 
smaller, thermal fluctuations play an increasingly relevant role.
Hence, this work acquires a stochastic contribution, i.e., the work
follows a distribution function $P(W)$. The shape of this function
depends on how the system is driven. If this change is induced by the
time variation of an external control parameter $\lambda(t)$, the 
distribution $P(W)$ becomes a functional of $\lambda(t)$.

Such distributions have recently become accessible experimentally for 
systems with only a few degrees of freedom diffusing in a thermal environment
under the influence of an externally controlled potential. Paradigmatic
examples include dragging a colloid particle by an optical tweezer through a 
viscous fluid~\cite{Wang02,Car04} and the forced unfolding of RNA
hairpins~\cite{Lip02}.
In both cases some realizations of the process show $W < \Delta F$.
Slightly overstated, such findings have been called violations of
the second law~\cite{Wang02}. In a more conservative interpretation of the
second law for such mesoscopic systems, the average work should and does still
obey
\begin{equation}
  \overline{W} \equiv \igl{W}{-\infty}{+\infty} P(W) W \geqslant \Delta F.
\end{equation}

Obviously, the distribution $P(W)$ is of paramount importance for a
better understanding of isothermal stochastic dynamics.
Exact statements, however, about $P(W)$ are scarce. In 1997, Jarzynksi 
has shown under rather mild assumptions that the distribution $P(W)$ obeys an
integral constraint
\begin{equation}
  \igl{W}{-\infty}{+\infty} P(W) e^{-\beta W} = e^{-\beta\Delta F}
  \label{eq:jarz}
\end{equation}
for any external protocol $\lambda(t)$~\cite{Jar97,Jar97-2,Cro00}. Here, $\beta
\equiv1/k_\text{B}T$ with Boltzmann's constant $k_\text{B}$. Since this
remarkable relation allows one to extract equilibrium free energy differences
from measuring or calculating the work distribution in non-equilibrium
experiments or simulations, it has found widespread applications
recently~\cite{Lip02,Humm01-2,Rit02,Park03,Sch04,ob}. The statistical and
convergence properties of this non-linear average deserve particular
attention~\cite{Zuck02,Gore03,Sh03}.

For time-dependent quadratic potentials, i.e. linear stochastic equations of 
motion, $P(W)$ can easily derived to be Gaussian~\cite{Maz99}.
Jarzynski's relation then implies that the mean $\overline{W}$ and the variance
$\sigma^2$ are necessarily related by~\cite{Jar97}
\begin{equation}
  \overline{W} = \Delta F + \beta\sigma^2/2.
  \label{eq:constraint}
\end{equation}
For these potentials, the Gaussian nature holds independently on the speed of
the driving, i.e. independently of how far the system is from equilibrium.

The purpose of this paper is to add a third general statement about
$P(W)$ to this list of exact results. We will show that this distribution
becomes a Gaussian for slow but finite driving even if the equations of motion
are non-linear. Since our approach is constructive, it yields an explicit
algorithm of how to obtain the mean $\overline{W}$ and the variance $\sigma^2$
of this Gaussian distribution. In the quasi-static limit of infinitely slow
external manipulation, this Gaussian reduces to $P(W)=\delta (W-\Delta F)$. 

A Gaussian character of the distribution $P(W)$ near equilibrium
seems to be expected or taken for granted in the recent literature
\cite{Jar97,Gore03,Humm01,Lip02,Sch04}.
Closer scrutiny of the references usually cited for this assumption -- if any
are cited at all --, however, reveals that they do not provide an explicit
proof of this statement. The often cited papers by Hermans~\cite{Her91} and
Wood et al.~\cite{Wood91} explicitly assume a Gaussian shape. Alluding in a
more general way to an Onsager-Machlup functional \cite{Groot} also fails,
since this Gaussian functional is derived for linear stochastic equations of
motion. The presumably most promising case to date in favor of a Gaussian
distribution suggests to invoke the central limit theorem for the increments
of work~\cite{Hen01}. However, without translating this proposal into a
definite calculation, which seems to be non-trivial for time-dependent
potentials, this argument is not a clear-cut proof yet, let alone does it give
an expression for mean and variance of this putative Gaussian.

Based on this unsatisfying state of affairs concerning such a fundamental
issue, we believe that a constructive derivation of the Gaussian nature of
this distribution for finite but slow driving is indeed called for as a step
towards a comprehensive theory of isothermal stochastic dynamics.

%%%%%%%%%%%%%%%%%%%%%%%%%%%%%%%%%%%%%%%%%%%%%%%%%%%%%%%%%%%%
%% Derivation
%%%%%%%%%%%%%%%%%%%%%%%%%%%%%%%%%%%%%%%%%%%%%%%%%%%%%%%%%%%%

% the system
For the derivation we consider a finite classical system coupled to a heat
reservoir of constant temperature. Let then $\vec{x}=(x_0,\dots,x_n)$ be the
state of the system with energy $V_\lambda(\vec{x})$ where $\lambda$ is an
externally controlled parameter. The stochastic dynamics is governed by the
Langevin equations~\cite{Gard}
\begin{equation}
  \dot{x}_i = -\mu_{ij}\pd{V_\lambda}{x_j}+\eta_i(t)
  \label{eq:lang}
\end{equation}
where $\mu_{ij}$ are the mobility coefficients~\footnote{For brevity we choose
constant mobility coefficients. Otherwise the Fokker-Planck operator
eq.~(\ref{eq:fp}) would contain spurious drift terms which would, however,
not affect our main result.} and $\eta_i(t)$ is the thermal noise representing
the heat bath with
\begin{equation}
  \mean{\eta_i(t)}=0 
  \text{~~and~~}
  \mean{\eta_i(t)\eta_j(t')}=\frac{2}{\beta}\mu_{ij}\delta(t-t'),
  \label{eq:markov}
\end{equation}
where $\mean{\dots}$ denotes the ensemble average.
We describe the continuous process of switching the system from an initial
state ($\lambda(t=0)=0$) to a final state ($\lambda(t=\ts)=1$) by a protocol
$\lambda(t)$, over a total switching time $\ts$. Without loss of generality,
we set
\begin{equation}
  \lambda(t) \equiv t/\ts
  \label{eq:lam}
\end{equation}
and hence a constant switching rate $\dlam=\ts^{-1}$~\cite{Jar97}.

% ensemble
We now consider an ensemble of infinitely many realizations of this Markov
process, each evolving stochastically according to eq.~(\ref{eq:lang}). The
normalized distribution of this ensemble in phase space $f(\vec{x},t)$ obeys a
Fokker-Planck equation~\cite{Gard}
\begin{equation}
  \partial_t f = \oL f 
  \text{~~with~~}
  \oL \equiv
      \pd{}{x_i}\mu_{ij} \left[ \pd{V_\lambda}{x_j}+\frac{1}{\beta}\pd{}{x_j}
      \right]
  \label{eq:fp}
\end{equation}
equivalent to the Langevin equations~(\ref{eq:lang}). This introduces the
(through $\lambda$ time-dependent) Fokker-Planck operator $\oL$. The
stationary solution of eq.~(\ref{eq:fp}) for fixed $\lambda$ is the
equilibrium distribution
\begin{equation}
  \feq(\vec{x}) \equiv e^{-\beta V_\lambda(\vec{x})} \big/
                       \ugl{\vec{x}'}e^{-\beta V_\lambda(\vec{x}')}.
  \label{eq:feq}
\end{equation}

% work
The total work performed along one particular trajectory $\vec{x}(t)$ up to
time $t$ is the time integral~\cite{Jar97,Sek98}
\begin{equation}
  W[\vec{x}(t),t] \equiv \igl{t'}{0}{t}\dlam\dV(\vec{x}(t')).
  \label{eq:work}
\end{equation}
We can now compose a combined stochastic process consisting of
$\{\vec{x},W\}$ as~\cite{Maz99}
\begin{eqnarray}
  &\dot{x}_i &= -\mu_{ij}\pd{V_\lambda}{x_j}+\eta_i(t) \\
  &\dot{W}   &= \dlam\dV.
  \label{eq:lang2}
\end{eqnarray}
Note that the equation of motion for $\dot{W}$ does not have an independent
noise but is stochastic through the $\vec{x}$-dependence of $V_\lambda$. The
joint probability distribution function $p(\vec{x},W,t)$ then obeys a
Fokker-Planck equation
\begin{equation}
  \partial_t p = \left[ \oL+\dlam\oW \right]p
  \label{eq:fp2}
\end{equation}
where
\begin{equation}
  \oW \equiv -\dV\pd{}{W}
\end{equation}
represents a drift term of the work. Eq.~(\ref{eq:fp2}) is exact, no matter how
far the system is driven out of equilibrium. The reduced probability
distribution of the work $P(W,t)$ can be obtained by integrating out $\vec{x}$
as
\begin{equation}
  P(W,t)=\ugl{\vec{x}}p(\vec{x},W,t).
\end{equation}
Since we start the process out of thermal equilibrium, the $\vec{x}$ are
initially distributed according to the canonical distribution and therefore
the initial condition is
\begin{equation}
  p(\vec{x},W,0) = f^\eq_0(\vec{x})\delta(W).
  \label{eq:initial}
\end{equation}

% projector
As our main theoretical tool we introduce a projector $\oP$ acting on a
function $\phi(\vec{x},W,t)$ such that
\begin{equation}
  \oP\phi \equiv \feq\ugl{\vec{x}'}\phi(\vec{x}',W,t).
  \label{eq:proj}
\end{equation}
Note that
\begin{equation}
  \oL\oP\phi = \oP\oL\phi = 0.
  \label{eq:cond2}
\end{equation}
The first statement is evident from definition~(\ref{eq:proj}) and the fact
that $\feq$ is in the null space of $\oL$. The second conclusion follows
when $\phi$ is expanded in terms of eigenfunctions of $\oL$. Then the
Fokker-Planck operator $\oL$ cancels the eigenfunction to eigenvalue 0, which
in fact is $\feq$, whereas the projector annihilates all other eigenfunctions
corresponding to higher eigenvalues.

% commutator of the projector
The other important property of the projector $\oP$, which distinguishes it
from the usual application to the adiabatic elimination of
fast variables~\cite{Gard}, is that it does not commute with the time
derivative but rather leads to
\begin{equation}
  [\partial_t,\oP] \equiv \partial_t\oP-\oP\partial_t
                   = -\dlam\beta S_\lambda\oP,
  \label{eq:comm}
\end{equation}
where we define
\begin{equation}
  S_\lambda \equiv \dV - \mean{\dV}_\lambda.
  \label{eq:S}
\end{equation}
The equilibrium ensemble average $\mean{\dots}_\lambda$ is defined as
\begin{equation}
  \mean{\phi}_\lambda \equiv \ugl{\vec{x}}\feq(\vec{x})\phi(\vec{x},t).
  \label{eq:eqmean}
\end{equation}

% expansion
We can now expand the joint probability $p(\vec{x},W,t)$ for small $\dlam$,
which corresponds to a separation of time scales~\cite{Kamp85}. The slow time
scale is $\lambda=t/\ts$. The fast time scale, which we do not need explicitly,
is determined by the intrinsic relaxation processes. As the time derivative
transforms according to $\partial_t\rightarrow\dlam\partial_\lambda$,
switching to the slow time scale eq.~(\ref{eq:fp2}) becomes
\begin{equation}
  \partial_\lambda p = \left[ \dlam^{-1}\oL+\oW \right]p.
  \label{eq:fp3}
\end{equation}
By using the projector $\oP$ we decompose the distribution function
$p=p_0+p_1$ into
\begin{equation}
  p_0(\vec{x},W,\lambda) \equiv \oP p = \feq(\vec{x})P(W,\lambda)
  \label{eq:p0}
\end{equation}
and
\begin{equation}
  p_1(\vec{x},W,\lambda) \equiv (1-\oP)p.
  \label{eq:p1}
\end{equation}
We apply $\oP$, respectively $(1-\oP)$, to eq.~(\ref{eq:fp3}) and keep in mind
both eq.~(\ref{eq:cond2}) and the commutator~(\ref{eq:comm}). We finally get
the two coupled differential equations
\begin{eqnarray}
  \label{eq:dgl:0}
  &\partial_\lambda p_0 &= \oA^0p_0+\oA^0p_1-\beta S_\lambda p_0 \\
  \label{eq:dgl:1}
  &\partial_\lambda p_1 &= \left[ \dlam^{-1}\oL+\oA^1 \right]p_1
                           +\oA^1p_0+\beta S_\lambda p_0
\end{eqnarray}
where we abbreviate
\begin{equation}
  \oA^0 \equiv \oP\oW \text{~~and~~} \oA^1 \equiv (1-\oP)\oW.
  \label{eq:A}
\end{equation}

% zero order
In this form, an expansion in $\dlam$ becomes possible. In lowest order
($\dlam\rightarrow 0$), eq.~(\ref{eq:dgl:1}) implies $\oL p_1=0$. Since $p_1$
is orthogonal to the null space of $\oL$ by definition~(\ref{eq:p1}), 
$p_1=0$ follows. For a solution of eq.~(\ref{eq:dgl:0}) we explicitly
calculate 
\begin{equation}
  \oA^0\feq = -\feq\ugl{\vec{x}}\dV\feq\pd{}{W} 
            = -\feq\mean{\dV}_\lambda\pd{}{W}.
\end{equation}
Using this and eq.~(\ref{eq:p0}) we finally obtain
\begin{equation}
  \pd{P}{\lambda} = -\mean{\dV}_\lambda\pd{P}{W}
\end{equation}
for the distribution of the work $P(W,\lambda)$. The solution of this equation
is $P(W)=\delta(W-\Delta F)$ with the initial condition $P(W,0)=\delta(W)$
following from eq.~(\ref{eq:initial}), where we recognize
\begin{equation}
  \Delta F = \igl{\lambda}{0}{1}\mean{\dV}_\lambda
\end{equation}
as the change in free energy of the entire process. We thus have recovered for
$\dlam\rightarrow 0$ the quasi-static limit as expected.

% first order
To first order in $\dlam$ we get from eq.~(\ref{eq:dgl:1})
\begin{equation}
  p_1 = -\dlam\oL^{-1}\left[ \oA^1+\beta S_\lambda \right]p_0.
\end{equation}
Putting this back into eq.~(\ref{eq:dgl:0}) we get after a straightforward
calculation a diffusion type equation for $P(W,\lambda)$ in the form
\begin{equation}
  \pd{P}{\lambda} = -\left[ \mean{\dV}_\lambda + \dlam\beta\tilde{S}_\lambda
  \right]\pd{P}{W} + \dlam\tilde{S}_\lambda\pd{^2P}{W^2}
  \label{eq:near}
\end{equation}
where
\begin{equation}
  \tilde{S}_\lambda \equiv -\ugl{\vec{x}}\dV\oL^{-1}S_\lambda\feq.
  \label{eq:SS}
\end{equation}
The solution is a Gaussian
\begin{equation}
  P(W) = \frac{1}{\sqrt{2\pi\sigma^2}}\exp\left[
         -\frac{(W-\overline{W})^2}{2\sigma^2} \right]
\end{equation}
with variance
\begin{equation}
  \sigma^2 = 2\dlam\igl{\lambda}{0}{1}\tilde{S}_\lambda
  \label{eq:var}
\end{equation}
and mean
\begin{equation}
  \overline{W} = \igl{\lambda}{0}{1}\left[ \mean{\dV}_\lambda +
  \dlam\beta\tilde{S}_\lambda \right] = \Delta F + \frac{\beta}{2}\sigma^2.
  \label{eq:mean}
\end{equation}
This is the central result of the present paper~\footnote{Since
$S_\lambda\feq$ has no component in the null space of $\oL$ ($\oP
S_\lambda\feq=0$), $\oL^{-1}S_\lambda\feq$ exists. A manifestly positive
definite expression for $\tilde{S}_\lambda$ can be obtained if the
distribution of the dissipated work, $P(W-\Delta F)$, is considered.}. First,
it proves that the distribution of the work in isothermal non-equilibrium
processes is Gaussian in the near-equilibrium regime. Second, we recover
independently from Jarzynski's relation~(\ref{eq:jarz}) the constraint that
the mean and variance are connected according to
eq.~(\ref{eq:constraint}). Third, it yields an explicit algorithm of how to
calculate these quantities.

For an assessment of the range of validity of this approximation, we recall
that it is based essentially on a separation of time scales. Hence, the
Gaussian distribution will be a good approximation as long as
\begin{equation}
  \dlam\tau \ll 1,
\end{equation}
where $\tau$ is an intrinsic relaxation time.

%%%%%%%%%%%%%%%%%%%%%%%%%%%%%%%%%%%%%%%%%%%%%%%%%%%%%%%%%%%%
%% Figures
%%%%%%%%%%%%%%%%%%%%%%%%%%%%%%%%%%%%%%%%%%%%%%%%%%%%%%%%%%%%
\begin{figure}[t]
  \includegraphics[height=5.67cm]{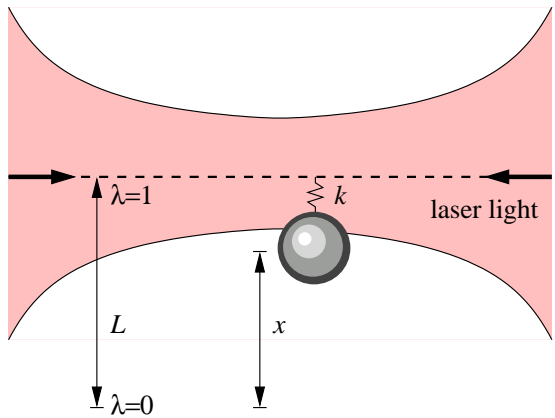}
  \caption{Scheme of the experiment of Ref.~\cite{Wang02}. A colloidal
    particle at position $x(t)$ is dragged by an optical tweezer with focus at
    $y(\lambda)=L\lambda$ through a viscous fluid. The effective potential is
    harmonic with a spring constant $k$.}
  \label{fig}
\end{figure}

%%%%%%%%%%%%%%%%%%%%%%%%%%%%%%%%%%%%%%%%%%%%%%%%%%%%%%%%%%%%
%% Illustration
%%%%%%%%%%%%%%%%%%%%%%%%%%%%%%%%%%%%%%%%%%%%%%%%%%%%%%%%%%%%

% example
As an example, we illustrate our approach for a simple one-dimensional case,
where we can compare our expansion with an exact solution~\cite{Maz99}. We
consider a colloidal particle at position $x$ with mobility $\mu$ trapped by
an optical tweezer whose center $y(\lambda)$ is moved at constant speed $v$
through a viscous fluid (see FIG.~\ref{fig}). The potential of the trap is
assumed to be harmonic near the focal point
\begin{equation}
  V_\lambda(x)=(k/2)(x-y(\lambda))^2
\end{equation}
with effective strength $k$. In this case, the free energy is independent of
$y(\lambda)$. For the two states, we choose with $y(\lambda=0)=0$ and
$y(\lambda=1)=L$ two positions of the trap. The switching rate becomes
$\dlam=v/L$, while the relaxation time is $\tau=1/\mu k$.

% calculating
Within our scheme, we first have to calculate $\Psi_\lambda \equiv
\oL^{-1}S_\lambda\feq$ in eq.~(\ref{eq:SS}) which amounts to solving the
inhomogeneous differential equation
\begin{equation}
  \oL\Psi_\lambda = S_\lambda\feq.
\end{equation}
With the Fokker-Planck operator from eq.~(\ref{eq:fp}) we get explicitly
\begin{multline}
  \mu\pd{}{x} \left[k(x-L\lambda)+\frac{1}{\beta}\pd{}{x}
  \right]\Psi_\lambda = \\
                        -\sqrt{\frac{\beta k^3}{2\pi}} L(x-L\lambda)
                        \exp\left[-\frac{1}{2}\beta k(x-L\lambda)^2\right]
  \label{eq:inv}
\end{multline}
where $\lambda$ only appears as a parameter. This is easily solved as
$\Psi_\lambda(x) = \feq(x)Lx/\mu$ and thus the average work becomes
\begin{equation}
  \overline{W} = L^2k\;\dlam\tau.
  \label{eq:work:app}
\end{equation}
Of course, for this harmonic potential, the distribution $P(W)$ is Gaussian at
any driving~\cite{Maz99}. The exact result for the mean $\overline{W}$
as a function of $\dlam$ reads
\begin{equation}
  \overline{W} = L^2k\left[ \dlam\tau-\dlam^2\tau^2\left( 1-e^{-1/\dlam\tau}
                 \right) \right],
  \label{eq:work:exact}
\end{equation}
which agrees to first order in $\dlam$ with eq.~(\ref{eq:work:app}) as
expected.

% singularity
The exact expression~(\ref{eq:work:exact}) points to an interesting property
which seems not to have been discussed yet in the context of stochastic
dynamics~\footnote{In quantum mechanics, exponentially small
terms as in eq.~(\ref{eq:work:exact}) occur for transition probabilities of
time-dependent two-state systems, see, e.g., M.~V.~Berry, Proc. R. Soc. A
{\bf 429}, 61 (1990).}. The exponentially small last term shows that the
average work is non-analytic in $\dlam$. We expect that if our expansion of
eq.~(\ref{eq:dgl:1}) was extended to the next order, some signature of this
non-analyticity should show up. Therefore the approach to equilibrium (or the
deviation from equilibrium) even in this almost trivial case is somewhat
subtle.

%%%%%%%%%%%%%%%%%%%%%%%%%%%%%%%%%%%%%%%%%%%%%%%%%%%%%%%%%%%%
%% Conclusions
%%%%%%%%%%%%%%%%%%%%%%%%%%%%%%%%%%%%%%%%%%%%%%%%%%%%%%%%%%%%

In summary, we have shown for general diffusive systems that the
distribution of work required to drive the system from an initial equilibrium
state to another is a Gaussian for slow but finite driving. Its mean and
variance can be obtained from solving an inhomogeneous differential equation
involving the Fokker-Planck operator. As an exactly solvable case shows,
these quantities are non-analytic in the switching rate. This result indicates
that in general calculating the next order correction to the Gaussian derived
here may face fundamental difficulties.

Stimulating discussions with O.~Braun and R.~Finken are gratefully
acknowledged as well as valuable hints and comments by H.~Spohn.

%%%%%%%%%%%%%%%%%%%%%%%%%%%%%%%%%%%%%%%%%%%%%%%%%%%%%%%%%%%%
%% Bibliography
%%%%%%%%%%%%%%%%%%%%%%%%%%%%%%%%%%%%%%%%%%%%%%%%%%%%%%%%%%%%

\end{document}